%% file: main.tex
\documentclass[]{article}

\usepackage{arxiv}

\usepackage[utf8]{inputenc} 
\usepackage[T1]{fontenc}    
\usepackage{url}            
\usepackage{booktabs}       
\usepackage{amsfonts}       
\usepackage{nicefrac}       
\usepackage{microtype}      
\usepackage{lipsum}
\usepackage{graphicx}
\usepackage{color,soul}
\usepackage{tikz}
\usepackage{tabularx} 
\usepackage{subcaption}
\usepackage{layouts}        
\usepackage{comment}        
\usepackage{siunitx}
\usepackage{todonotes}
\usepackage{placeins}

\usepackage{nth}
\usepackage[acronym]{glossaries} 

\newlength{\twosubht}
\newsavebox{\twosubbox}
\newcommand{\NineEuroTicket}{\mbox{9\ EUR-Ticket}}

\newacronym{pt}{PT}{public transport}
\newacronym{mid}{MiD}{Mobilität in Deutschland}

\title{A nation-wide experiment: fuel tax cuts and almost free public transport for three months in Germany - Report 5  Insights into four months of mobility tracking}

\author{
Lennart Adenaw\\
Technical University of Munich\\
TUM School of Engineering and Design\\
Chair of Automotive Technology \\
Boltzmannstrasse 15, 85748 Garching\\
\texttt{lennart.adenaw@tum.de}\\
\And
David Ziegler\\
Technical University of Munich\\
TUM School of Engineering and Design\\
Chair of Automotive Technology \\
Boltzmannstrasse 15, 85748 Garching\\
\texttt{david.ziegler@tum.de}\\
\And 
Nico Nachtigall\\
Technical University of Munich\\
TUM School of Engineering and Design\\
Chair of Automotive Technology \\
Boltzmannstrasse 15, 85748 Garching\\
\texttt{nico.nachtigall@tum.de}\\
\And 
Felix Gotzler\\
Technical University of Munich\\
TUM School of Engineering and Design\\
Chair of Automotive Technology \\
Boltzmannstrasse 15, 85748 Garching\\
\texttt{felix.gotzler@tum.de}\\
\And
Allister Loder \\
Technical University of Munich\\
TUM School of Engineering and Design\\
Chair of Traffic Engineering and Control\\
Arcisstrasse 21, 80333 Munich \\
\texttt{allister.loder@tum.de}\\
\And
Markus B. Siewert\\
Munich School of Politics and Public Policy\\
TUM Think Tank\\
Richard-Wagner-Straße 1, 80333 München\\
\texttt{markus.siewert@hfp.tum.de}
\And
Markus Lienkamp\\
Technical University of Munich\\
TUM School of Engineering and Design\\
Chair of Automotive Technology \\
Boltzmannstrasse 15, 85748 Garching\\
\texttt{lienkamp@tum.de}\\
\And
Klaus Bogenberger\\
Technical University of Munich\\
TUM School of Engineering and Design\\
Chair of Traffic Engineering and Control\\
Arcisstrasse 21, 80333 Munich\\
\texttt{klaus.bogenberger@tum.de}\\
}

\begin{document}
\maketitle


\begin{abstract}
In spring 2022, the German federal government agreed on a set of measures that aim at reducing households' financial burden resulting from a recent price increase, especially in energy and mobility. These measures include among others, a \mbox{nation-wide} public transport ticket for 9 EUR per month and a fuel tax cut that reduces fuel prices by more than \SI{15}{\percent}. In transportation research this is an almost unprecedented behavioral experiment. It allows to study not only behavioral responses in mode choice and induced demand but also to assess the effectiveness of transport policy instruments. We observe this natural experiment with a \mbox{three-wave} survey and an \mbox{app-based} travel diary on a sample of hundreds of participants as well as an analysis of traffic counts. In this fifth report, we present first analyses of the recorded tracking data. 910 participants completed the tracking until September, 30th. First, an overview over the \mbox{socio-demographic} characteristics of the participants within our tracking sample is given. We observe an adequate representation of female and male participants, a slight \mbox{over-representation} of young participants, and an income distribution similar to the one known from the "Mobilit\"at in Deutschland" survey \cite{BundesministeriumfurVerkehrunddigitaleInfrastruktur.2018}. Most participants of the tracking study live in Munich, Germany. General transportation statistics are derived from the data for all phases of the natural experiment \textendash\ prior, during, and after the \NineEuroTicket\ \textendash\ to assess potential changes in the participants' travel behavior on an aggregated level. A significant impact of the \NineEuroTicket\ on modal shares can be seen. An analysis of the participants' mobility behavior considering trip purposes, age, and income sheds light on how the \NineEuroTicket\ impacts different social groups and activities. We find that age, income, and trip purpose significantly influence the impact of the \NineEuroTicket\ on the observed modal split.
\end{abstract}


\input{_sections/1_introduction}
\input{_sections/2_tracking_participants}
\input{_sections/3_transport_statistics}
\input{_sections/4_mobility_behavior_analysis}
\input{_sections/5_discussion}


\section*{Acknowledgements}

The authors would like to thank the TUM Think Tank at the Munich School of Politics and Public Policy led by Urs Gasser for their financial and organizational support and the TUM Board of Management for supporting personally the genesis of the project. The authors thank the company MOTIONTAG for their efforts in producing the app at unprecedented speed. Further, the authors would like thank everyone who supported us in recruiting participants, especially Oliver May-Beckmann and Ulrich Meyer from M Cube and TUM, respectively. This project is partially funded by the Bavarian State Ministry of Science and the Arts in the framework of the bidt Graduate Center for Postdocs. 


\bibliography{references}  


\end{document}

%% file: _sections/1_introduction.tex
\section{Introduction}
\label{sec:introduction}

In transportation research, it is quite unlikely to observe or even perform real-world experiments in terms of travel behavior or traffic flow. There are few notable exceptions: subway strikes suddenly make one important alternative mode not available anymore  \cite{Anderson2014,Adler2016}, a global pandemic changes travelers' preferences for traveling at all or traveling collectively with others \cite{Molloy2021}, or a bridge collapse forces travelers to alter their daily activities \cite{Zhu2010}. However, in 2022 the German federal government announced in response to a sharp increase in energy and consumer prices a set of measures that partially offset the cost increases for households. Among these are a public transport ticket at 9\ EUR per month\footnote{\url{https://www.bundesregierung.de/breg-de/aktuelles/9-euro-ticket-2028756}} for traveling all across Germany in public transport, except for long-distance train services (e.g., ICE, TGV, night trains), as well as a tax cut on gasoline and diesel, resulting in a cost reduction of about 15\ \% for car drivers\footnote{\url{https://www.bundesfinanzministerium.de/Content/DE/Standardartikel/Themen/Schlaglichter/Entlastungen/schnelle-spuerbare-entlastungen.html}}. Both measures are limited to three months, namely June, July and August 2022. As of mid June, more than 16\ million tickets have been sold\footnote{
\url{https://www.tagesschau.de/wirtschaft/unternehmen/neun-euro-ticket-135.html}
}, while it seems that the fuel tax cut did not reach consumers due to generally increased fuel prices and oil companies are accused of not forwarding the tax cuts to consumers \footnote{\url{https://apnews.com/article/politics-business-germany-prices-deb85a000d63cd57b76446d9c90c3e18}}. 

For the Munich metropolitan region, Germany, we designed a study comprising three elements. The three elements are: (i) a three-wave survey before, during and after the introduction of cost-saving measures; (ii) a smartphone app based measurement of travel behavior and activities during the same period; (iii) an analysis of aggregated traffic counts and mobility indicators. We will use data from 2017 (pre-COVID-19) and data from shortly before the cost reduction measures as the control group. In addition, the three-wave survey is presented to a German representative sample. The main goal of the study is to investigate the effectiveness of the cost-saving measures with focus on the behavioral impact of the \NineEuroTicket\ on mode choice \cite{ben1985discrete}, rebound effects \cite{Greening2000,Hymel2010}, and induced demand \cite{Weis2009}. Further details on the study design and the first results can be found in our four previous reports \cite{reportone, reporttwo, report3, report4}.

In this fifth report, we first provide an assessment of the sociodemographics of the sample group actively using our smartphone tracking app in \mbox{Section \ref{sec:tracking_participants}}. In \mbox{Section \ref{sec:aggregated_statistics}}, we supply aggregated transportation statistics from the tracking data that has been recorded from beginning of the experiment to end of September 2022. \mbox{Section \ref{sec:data_analysis}} analyzes the mobility of the participants regarding their \mbox{spatio-temporal} activity behavior taking into account the different \mbox{socio-demographic} groups among our participants. The results of the exploratory data analysis presented herein, are discussed in \mbox{Section \ref{sec:discussion}}.

%% file: _sections/2_tracking_participants.tex
\section{Tracking Participants}
\label{sec:tracking_participants}

This section elaborates on the \mbox{socio-economic} characteristics of the 910 \mbox{GPS-tracked} participants who took part in the study from \mbox{May 24, 2022} via the ``Mobilität.Leben'' smartphone application available on iOS and Android devices and recorded at least one trip until \mbox{September 30, 2022}. All \mbox{socio-economic} parameters were \mbox{self-reported} by the participants within the survey phase of the study. For general results of this phase, the interested reader is referred to our previous report \cite{reporttwo}. This evaluation especially focuses on the tracked participants, constituting a subsample of all surveyed participants.

\newpage

\subsection{Age Distribution of Participants}
\begin{figure}[!htbp]%
	\centering%
	\includegraphics[width=.7\textwidth]{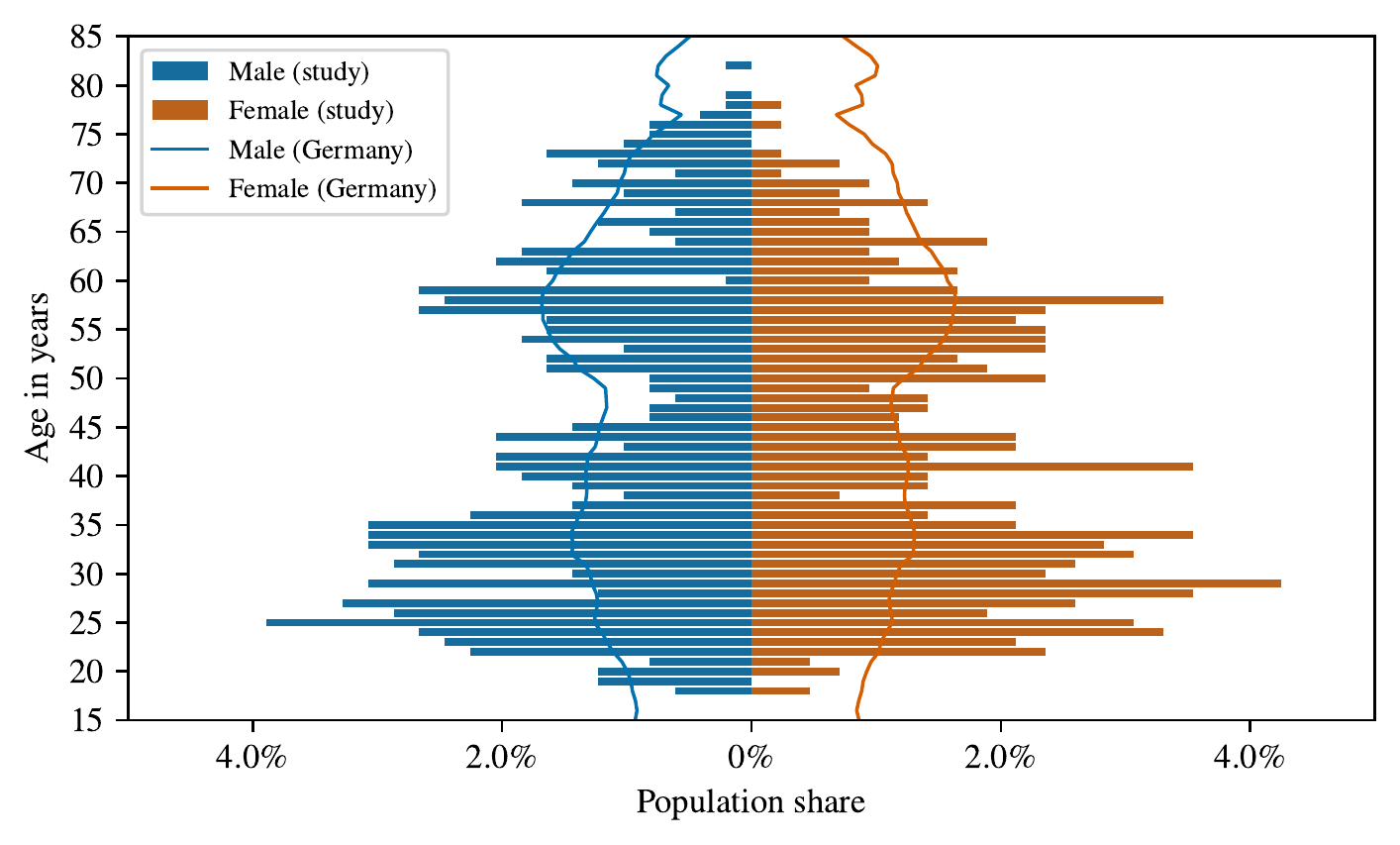}
	\caption{Age distribution of tracked participants vs. age distribution for Germany in 2022}%
	\label{fig:age_distribution}%
\end{figure}

\mbox{Figure \ref{fig:age_distribution}} shows the relative age distribution of male and female participants within our tracking sample. For reference, the demographic age distribution for Germany in 2022 is depicted. Comparing the age distribution of our study group with the German age distribution, the following observations can be made: for male participants between 40 to 80 years, the participant distribution aligns well with the German age distribution, whereas participants of both genders with an age between 20 and 40 are \mbox{over-represented}. Women between the age of 65 and 80 are \mbox{under-represented} in comparison to men in the same age group. Participants younger than 18 do not occur as a minimum age of 18 is required to participate in the study. Due to the general availability of data within all relevant age groups, we conclude that with appropriate sample weights we can make our tracking sample representative from an age and gender perspective in future analyses. Since this report is focused on a descriptive analysis of the recorded data, we leave an extrapolation using such weights to later research.

\FloatBarrier

\subsection{Gender Distribution of Participants}
\begin{figure}[!htbp]%
	\centering%
	\includegraphics[width=.6\textwidth]{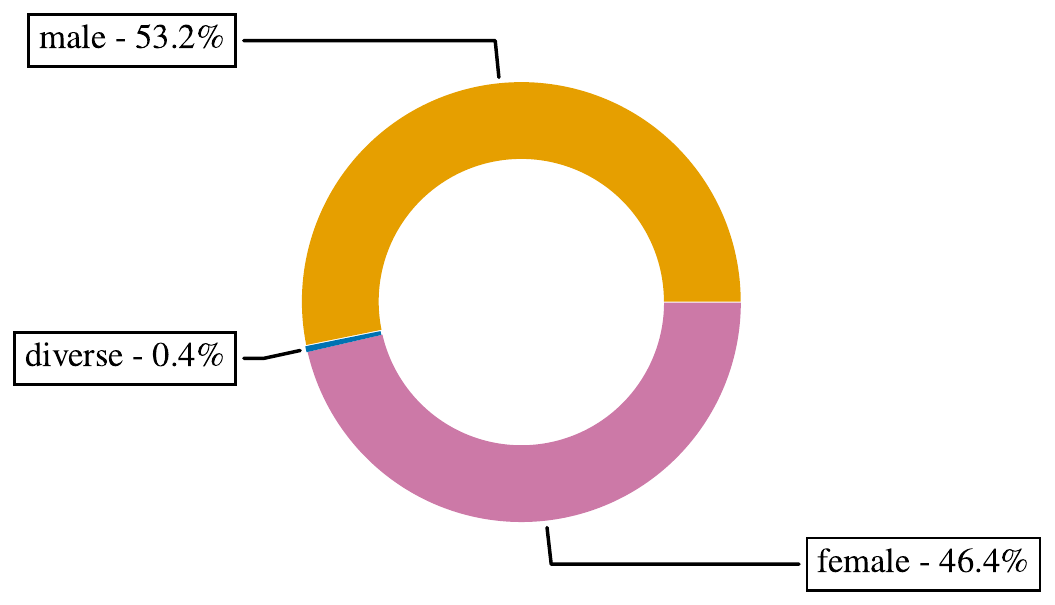}
	\caption{Gender distribution of tracked participants}%
	\label{fig:gender_distribution}%
\end{figure}

We asked participants to report their gender as one of three categories: \textit{male}, \textit{female}, and \textit{diverse}. The sample contains approximately equal numbers of  female and male participants (422 female, 484 male and 4 diverse), as shown in \mbox{Figure \ref{fig:gender_distribution}}.

\FloatBarrier

\subsection{Income of Participants}
\begin{figure}[!htbp]%
	\centering%
	\includegraphics[width=.85\textwidth]{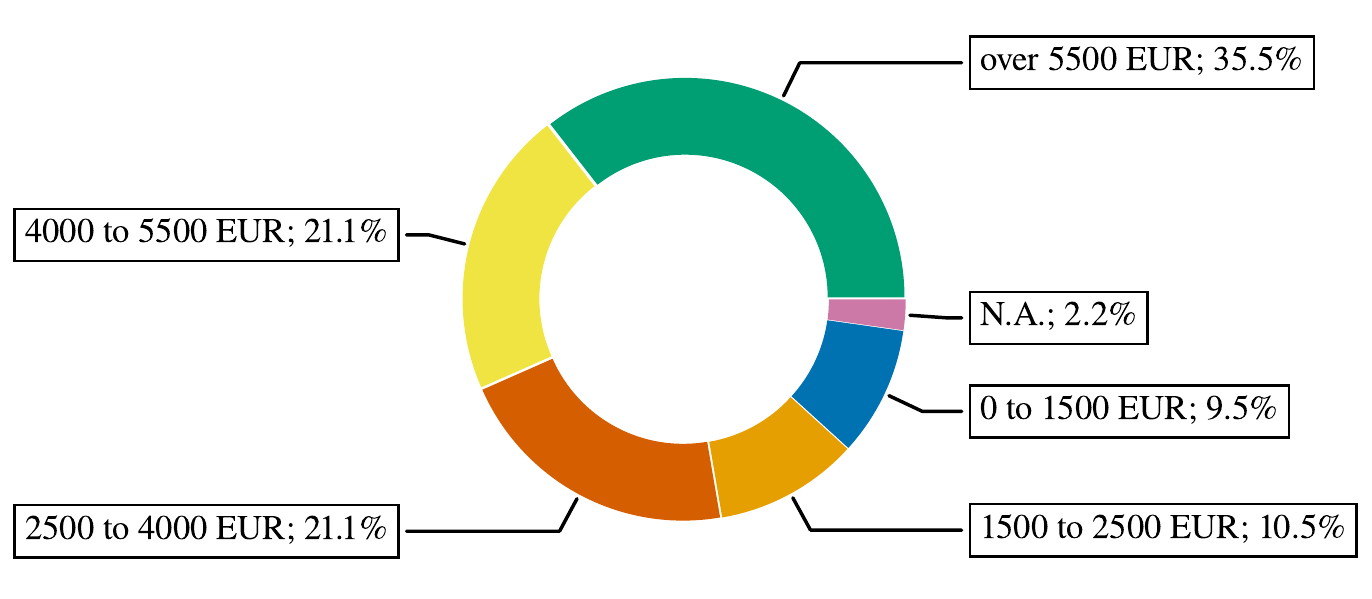}
	\caption{Income of tracked participants}%
	\label{fig:income_distribution}%
\end{figure}

We evaluated participants' income levels using threshold definitions already known from the \gls{mid} \cite{BundesministeriumfurVerkehrunddigitaleInfrastruktur.2018} study, as shown in \mbox{Figure \ref{fig:income_distribution}}. The evaluation shows that with a share of \SI{35.5}{\percent} most households have a net income of more than \mbox{5,500~€} per month, followed by two groups of \SI{21.2}{\percent} each, of households with incomes between \mbox{4,000~€} and \mbox{5,500~€} and between \mbox{2,500~€} and \mbox{4,000~€}, respectively. \SI{10.5}{\percent} of the sample households have between \mbox{1,500~€} and \mbox{2,500~€} at their monthly disposal. The number of participants in lower household income groups is significantly smaller with \SI{9.5}{\percent} living on less than \mbox{1,500~€} per month. \SI{2.2}{\percent} did not disclose their income.

\FloatBarrier

\subsection{Dwelling}
\begin{figure}[htp]

\sbox\twosubbox{%
  \resizebox{\dimexpr.99\textwidth-1em}{!}{%
    \includegraphics[height=3cm,]{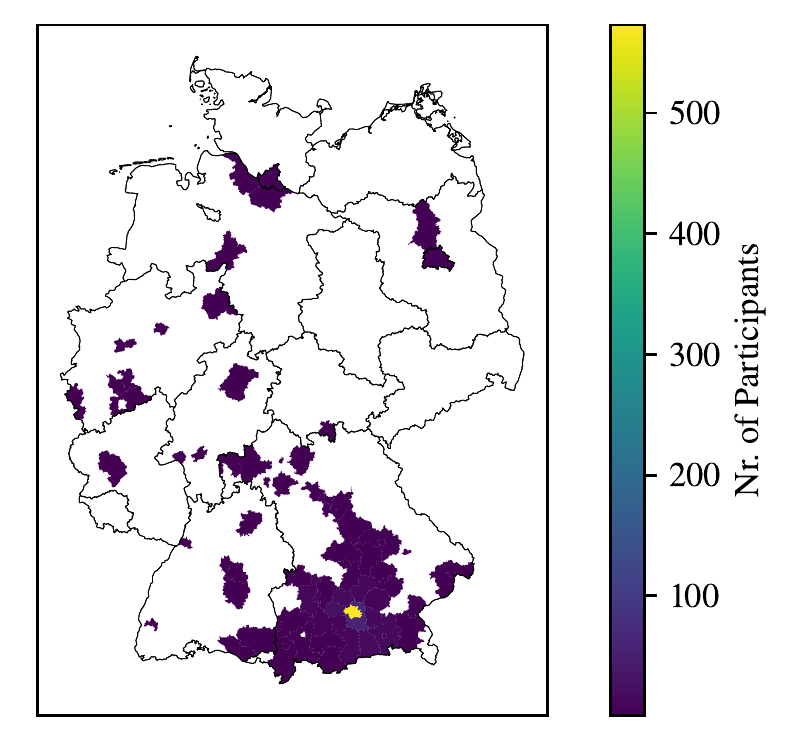}%
    \includegraphics[height=3cm]{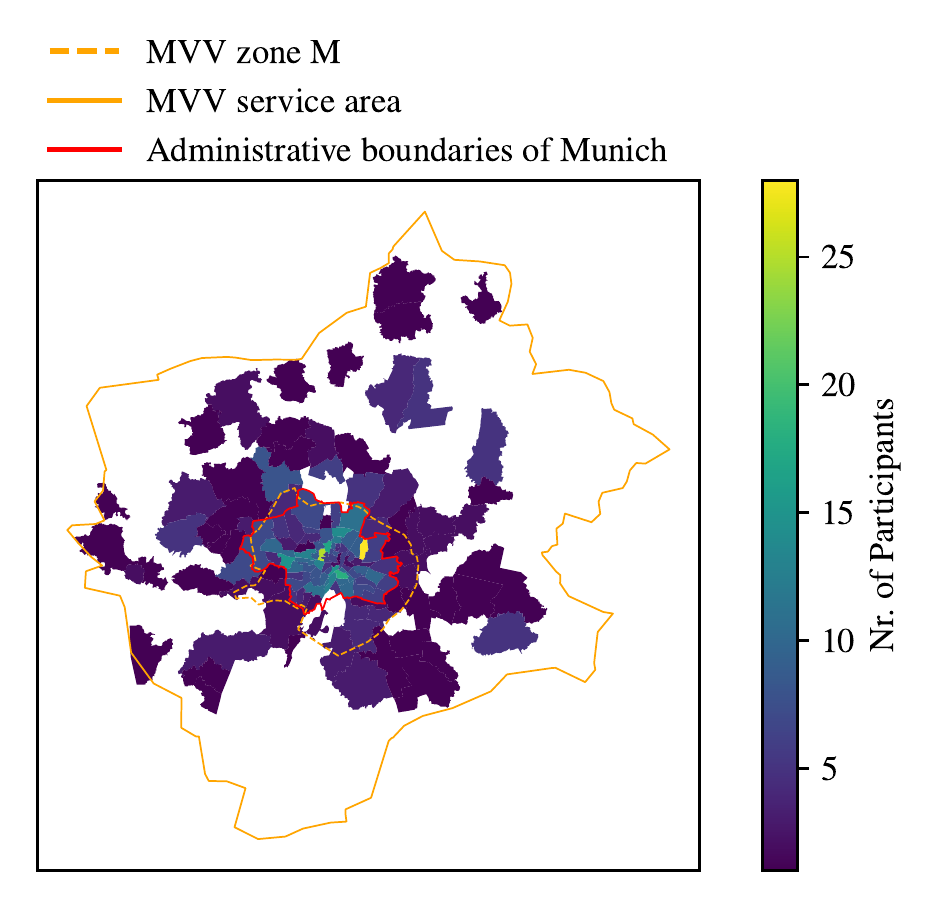}%
  }%
}
\setlength{\twosubht}{\ht\twosubbox}

\centering
\subcaptionbox{Complete sample group\label{fig:dwelling_map_de}}{%
  \includegraphics[height=\twosubht]{_figure/participant_map.pdf}%
}\quad
\subcaptionbox{Study participants in Munich area \label{fig:dwelling_map_muc}}{%
  \includegraphics[height=\twosubht]{_figure/participant_map_by.pdf}%
}
\caption{Residences of sample group}
\end{figure}

Participants of the tracking experiment mainly live in the south east of Germany, where the study was first announced, as shown in \mbox{Figure \ref{fig:dwelling_map_de}}. A majority resides in the greater Munich area and most within the city of Munich and its central and thus most interconnected \gls{pt} tariff-zone (zone \textit{M}) of the local \gls{pt} agency \mbox{MVV}. (\mbox{Figure \ref{fig:dwelling_map_muc}}). Concludingly, \gls{pt} is easily available to the tracking participants.

%% file: _sections/3_transport_statistics.tex
\section{Transport Statistics}
\label{sec:aggregated_statistics}

\begin{figure}[!htbp]%
	\centering%
	\includegraphics[width=\textwidth]{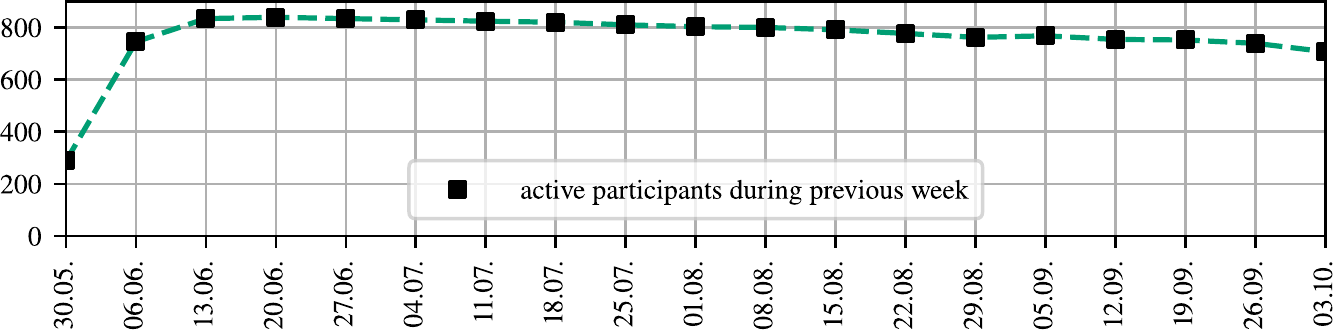}%
	\caption{App usage: Number of participants who recorded at least one trip during the previous week}%
	\label{fig:participants}%
\end{figure}

In this section, an overview over aggregated transport statistics (traveled distances, modal split) is presented together with an update on participation rates. All analyses are performed based on tracking data recorded between \mbox{May 24,\ 2022} and \mbox{September 30,\ 2022}. This includes the total time range from beginning of June until the end of August during which the \NineEuroTicket\ was valid in Germany.
During this observation period, \SI{910}{} individuals have recorded a trip with the provided smartphone application at least once. In \mbox{Figure \ref{fig:participants}} the amount of active participants per week can be seen. In it, a participant is considered to be active in a given week if they recorded at least one trip in that week. It can be seen that more than \SI{700}{} participants have been active each week after the first week of the study. During the first week of the experiment, around \SI{300}{} participants were recorded. The first week of our experiment was also the last week before the \NineEuroTicket\ was introduced. After a high increase of active participants in the second and third week participation peaked in the fourth week with around \SI{820}{} active users. During the following weeks, the amount of participants slowly decreased with \SI{700}{} active participants in the last week of our observation period.\\

A total of \SI{670096}{} trips was recorded within the observation period assessed in this report. Based on participants' feedback via the smartphone application, \SI{4982}{} trips were recorded incorrectly due to technical errors. The remaining \SI{665114}{} trips have a total length of \SI{6550250}{\kilo\meter}. Of these, walking was the most prominent mode of transport with \SI{372430}{} reported trips, followed by individual transport, e.g. car, with \SI{108944}{} trips and \SI{101265}{} trips by public transport. Trips by bike were recorded in \SI{80088}{} cases and by airplane and all other forms of transport in \SI{786}{} and \SI{1601}{} respectively. While the share of trips conducted by plane is low, accounting for only \SI{0.1}{\percent} of all recorded trips, distances travelled by plane account for \SI{21.2}{\percent} of the recorded mileage. Due to this disbalance, we exclude air travel from the following analysis to not skew our results. We also disregard all recorded trips with \mbox{non-specified} forms of transport. Our used smartphone application can accept feedback from users. They can correct the detected mode of transport, mark a trip as incorrect, or merge two trips as one. These corrections were made for \SI{36151}{} trips. It is also possible to confirm the recorded trips. This has been done by our participants for \SI{430199}{} trips.

\begin{figure}[!htbp]%
	\centering%
	\includegraphics[width=\textwidth]{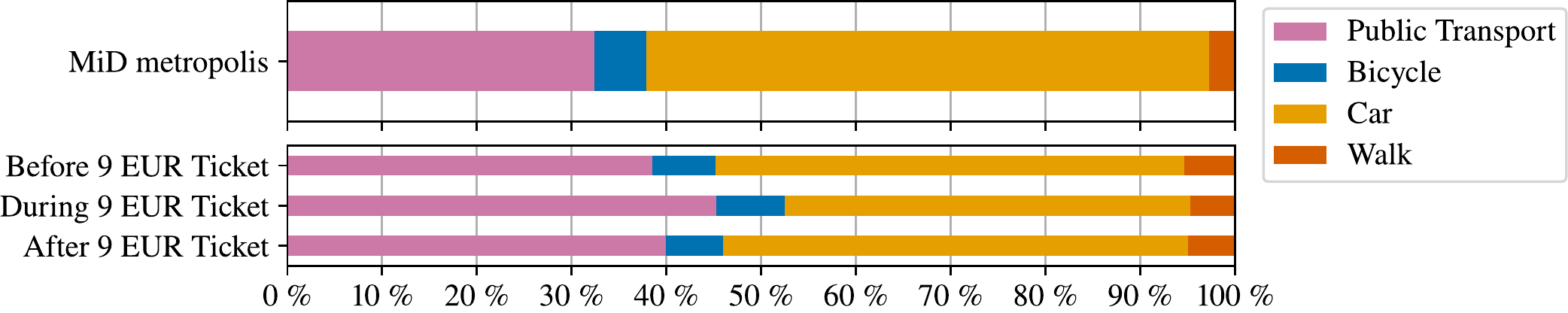}%
	\caption{Modal share based on the traveled distance. From top to bottom: data derived from \gls{mid} \cite{BundesministeriumfurVerkehrunddigitaleInfrastruktur.2018} for metropolis regions; recorded data with our smartphone app before, during, and after the \NineEuroTicket\ period}%
	\label{fig:ModalShare}%
\end{figure}

To assess the impact of the \NineEuroTicket\ on the general travel behavior of our sample we divided the reported data into three parts: the first part includes all trips recorded before the \NineEuroTicket\ was introduced on June\ 1,\ 2022. The second part contains all recorded trips between June\ 1,\ 2022 and August\ 31,\ 2022. In this time period the \NineEuroTicket\ was valid. The last and third part includes all trips between September\ 1 and September\ 30,\ 2022 and represents the time period directly after the \NineEuroTicket. \mbox{Figure \ref{fig:ModalShare}} reveals the modal share of our sample group in terms of traveled distances. Herein, modal shares are depicted for all phases of the experiment. For reference, modal shares for metropolitan regions as taken from the \gls{mid} survey are shown. To allow for a more concise presentation, we aggregate the means of transport as implemented in the smartphone app in the following manner:

\begin{itemize}
    \item \emph{Public Transport}: subway, light rail, regional train, train, bus, tram
    \item \emph{Bicycle}: bicycle, bikesharing, e-bicycle, kickscooter
    \item \emph{Car}: car, e-car, motorbike, taxi, uber, carsharing
    \item \emph{Walk}: walk
\end{itemize}

A comparison of the 2017 modal shares from the \gls{mid} study to our data reveals principle differences between the data sets. While \gls{mid} reports \SI{32.4}{\percent} public transport, \SI{5.4}{\percent} bicycle, and \SI{59.5}{\percent} car use while indicating \SI{2.7}{\percent} of walking, we observe a higher use of public transport and a lower car use within all phases of the experiment. 
Setting the modal shares in the three phases of our experiment side by side, it can be seen that the introduction of the \NineEuroTicket\ had no clear impact on the share of distance covered by bicycle. This remained almost constant at \SI{6.7}{\percent} percent and \SI{7.2}{\percent} percent. After the \NineEuroTicket\ program ended, the share decreased to \SI{5.9}{\percent}, most likely due to decreasing temperatures (the mean temperature in Munich was \SI{20.2}{\celsius} in the time range before and during the \NineEuroTicket\ and decreased to \SI{14.2}{\celsius} afterwards)\footnote{Calculated based on the public available LMU weather data for Munich \url{https://www.meteo.physik.uni-muenchen.de/DokuWiki/doku.php?id=wetter:stadt:messung}}. 
In contrast, the recorded share of walking clearly decreased from \SI{5.4}{\percent} of the total recorded mileage before to \SI{4.7}{\percent} during the \NineEuroTicket\, albeit almost constant weather conditions. This could indicate that the hurdle to use public transport for smaller distances diminished for our sample group with the cheap ticket.\\
The observed share of individual transport and public transport usage behaves the other way around: before the \NineEuroTicket\ was introduced, the share of individual transport was \SI{49.4}{\percent}. It then decreased by \SI{6.6}{\percent} during the time the ticket was valid. In the observed \mbox{time-frame} after the \NineEuroTicket\ the distance share of individual transport increased by \SI{6.2}{\percent} to a total share of \SI{49.0}{\percent}; slightly less than before the \NineEuroTicket.\
Initially, our sample group used public transport for \SI{38.5}{\percent} of the recorded distance. This share increased by \SI{6.8}{\percent} to \SI{45.3}{\percent} with the \NineEuroTicket\ before it decreased to \SI{40.0}{\percent} when the \NineEuroTicket\ program ended.\\

\begin{figure}[!htbp]%
	\centering%
	\includegraphics[width=\textwidth]{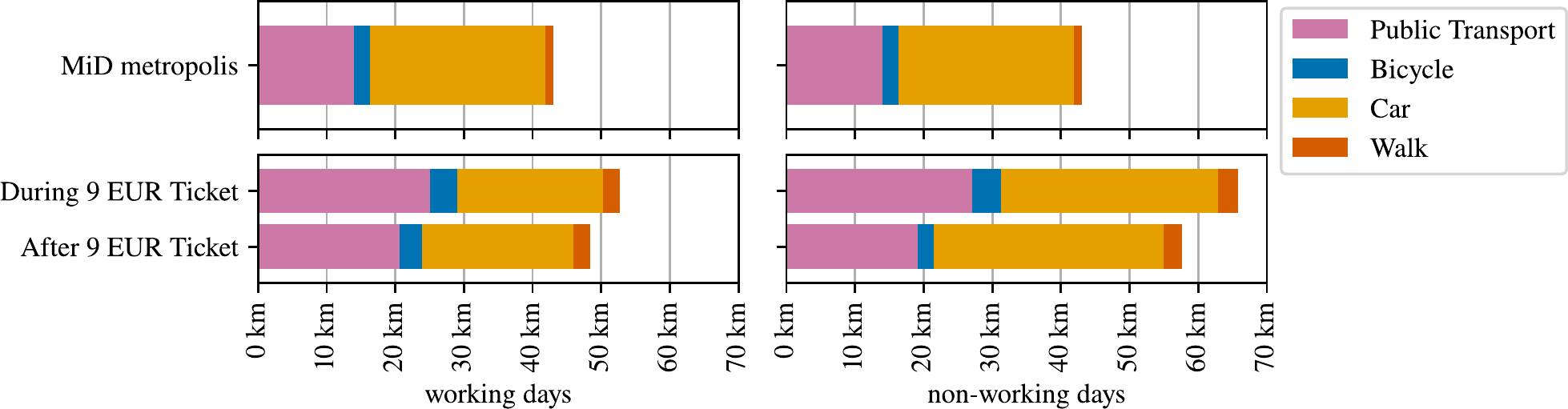}%
	\caption{Average traveled distance in \si{\kilo\meter} per active user, transport mode and day}%
	\label{fig:Distance}%
\end{figure}

In \mbox{Figure \ref{fig:Distance}} the total travel distance per day and mode of transport for an average person from our sample group is shown for working days and \mbox{non-working} days comprising Saturdays, Sundays, and public holidays. Note that from hereon no data from before the \NineEuroTicket\ is shown, because the relatively small sample size in this data set caused by a smaller number of participants and a shorter observation \mbox{time-frame} appears to be unsuited for lower level disaggregations. Instead, we compare the results of our sample group with results from the \gls{mid} study.\\
Overall it can be seen that if a person of our sample group is active on a \mbox{non-working} day the average travel distance with all modes of transport is equal or higher than on an average day during the week. The only two exceptions are the bicycle and the public transport usage in the period after the \NineEuroTicket.
The walking distance remained almost constant over the two time periods. During the week our participants walked \SI{2.5}{\kilo\meter} and on \mbox{non-working} days between \SI{2.8}{\kilo\meter} and \SI{2.9}{\kilo\meter}. It is noteworthy that \gls{mid} reports significantly smaller walking distances, probably due to the fact that people were not tracked but surveyed. 
The distance our sample group traveled by public transport decreased clearly after the \NineEuroTicket\ period. On an average working day by \SI{4.4}{\kilo\meter} and even more on a \mbox{non-working} day on average by \SI{7.9}{\kilo\meter}. Nevertheless, the distance traveled by individual transport did not increase by the same amount. We recorded an increase on working days by \SI{0.9}{\kilo\meter} and on \mbox{non-working} days by \SI{1.9}{\kilo\meter}. This means that the total distance traveled by an average participant decreased especially on \mbox{non-working} days. After the \NineEuroTicket\ ended on \mbox{August 31, 2022} the distance traveled by public transport decreased to \SI{20.7}{\kilo\meter} on working days and to \SI{19.1}{\kilo\meter} on \mbox{non-working} days. 
The distance traveled by means of individual transport increased to \SI{22.1}{\kilo\meter} on working days. On \mbox{non-working} days the traveled distance by individual transport increased to \SI{33.5}{\kilo\meter} per user and day. 
The total recorded distance per user and day amounts to \SI{52.7}{\kilo\meter} during and \SI{48.4}{\kilo\meter} after the \NineEuroTicket\ on working days and \SI{65.8}{\kilo\meter} and \SI{57.7}{\kilo\meter} on \mbox{non-working} days. In contrast, \gls{mid} reports an average travel distance of \SI{37}{\kilo\meter} per day and person as well as lower average distances for each individual mode. This difference is partly based in the fact that \gls{mid} values were determined for all participants including those who did not report any trip on a given day. Since our data considers only the mobile part of our sample population, \mbox{i.e.} persons that moved at least once during the day, we adjusted the \gls{mid} reference values in \mbox{Figure \ref{fig:Distance}} based on the \gls{mid} mobility rate. Nevertheless, a significant difference remains in all categories as well as in total mileage. The reasons for this gap may be subject to future research.

%% file: _sections/4_mobility_behavior_analysis.tex
\section{Mobility Behavior Analysis}
\label{sec:data_analysis}

In the subsequent section of this report, we investigate the mobility behaviour of the sample group. Thereby, we focus on the effects the \NineEuroTicket\ on the modal shares of different demographic groups, described by age and income, as well as varying trip purposes. In contrast to previous sections, we do not analyze the period before the introduction of the \NineEuroTicket\ in this section. This is due to the further subdivision of trips within the phases, e.g. by trip purposes, which results in significant differences in remaining total trip quantities and finally would lead to very small sample sizes for certain segments within the phase before the \NineEuroTicket. Instead, we use the \gls{mid} \cite{BundesministeriumfurVerkehrunddigitaleInfrastruktur.2018} as a reference and compare the results of our sample group to the national average for German metropolises. We do only consider trips conducted within the geographic borders of Germany.\\
In our analyses of the modal splits, we calculate the respective shares based on traveled distance by respective modes. We apply the concept of main trip purposes (\emph{Hauptwegezwecke}) from \gls{mid}, which allocates the complete distance of an intermodal trip to the means of transport with the highest share of the total trip length. In order to facilitate concise graphic representations of the results, we aggregate trip purposes available in the smartphone application in the following way:
\begin{itemize}
    \item \emph{home}: home
    \item \emph{work/education}: study, work
    \item \emph{leisure}: sport, eat, family and friends, leisure
    \item \emph{errand}: assistance, medical visit, errand
    \item \emph{shopping}: shopping
    \item \emph{unknown}: unknown, other
\end{itemize}

For the same reason, we aggregate means of transport in the manner described in \mbox{Section \ref{sec:aggregated_statistics}}. Due to the relatively small share of walking trips in respect to the total distance and for the sake of clarity and simplicity, this mode is not considered in the subsequent analyses. Values are normalized to the total distance covered with the remaining three means of transport. In order to account for the known differences in mobility behavior between working and \mbox{non-working} days \cite{Gerike.2018}, we strictly distinguish between said working days and weekends and public holidays (once more grouped as \mbox{non-working days}).

\begin{figure}[htp]
\sbox\twosubbox{%
  \resizebox{\dimexpr.99\textwidth-1em}{!}{%
    \includegraphics[height=3cm]{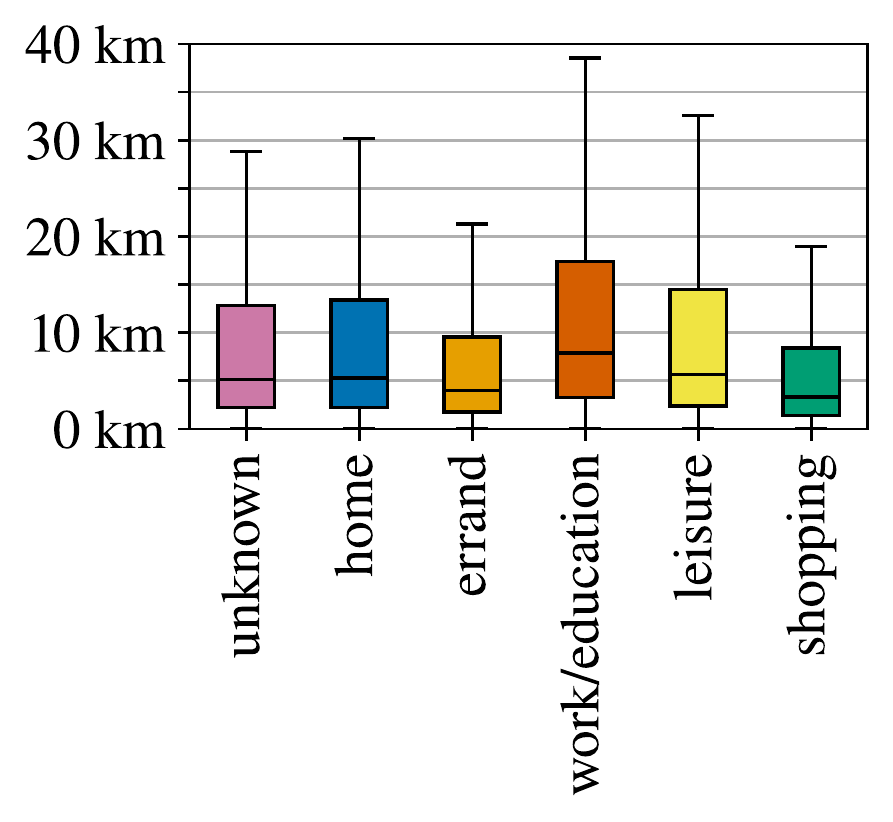}%
    \includegraphics[height=3cm]{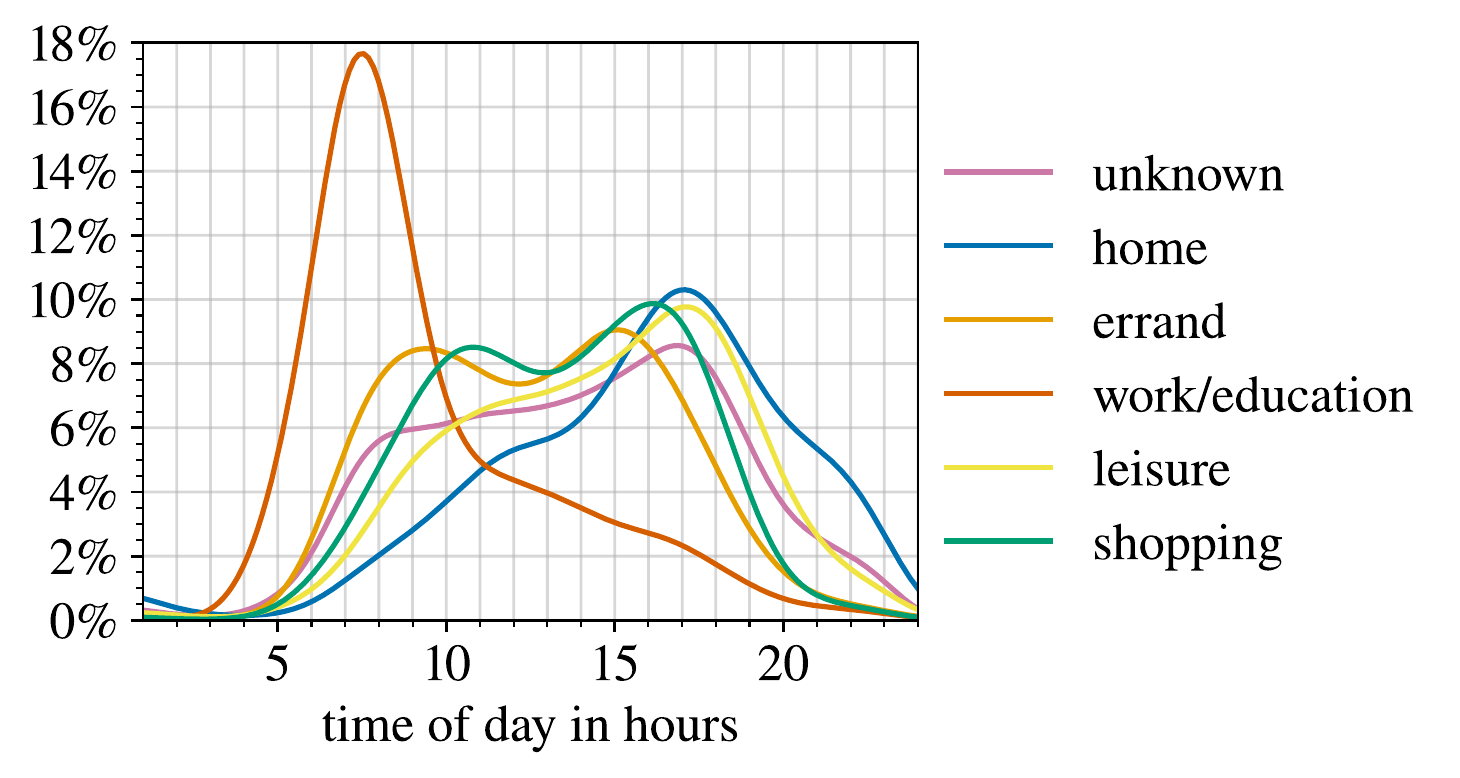}%
  }%
}
\setlength{\twosubht}{\ht\twosubbox}
\centering
\subcaptionbox{distances of trips by purpose\label{fig:MS_overview_dist}}{%
  \includegraphics[height=\twosubht]{_figure/dist_trips_purp_svg-tex.pdf}%
}\quad
\subcaptionbox{temporal distribution of trips by purpose\label{fig:MS_overview_start}}{%
  \includegraphics[height=\twosubht]{_figure/starttimes_svg-tex.pdf}%
}
\caption{General analysis of different trip purposes}
\end{figure}

An overview over the general mobility behavior of the tracking group is supplied in \mbox{Figures \ref{fig:MS_overview_dist} and \ref{fig:MS_overview_start}}. The purpose was either automatically inferred by the smartphone application based on e.g. points of interest or manually assigned by participants. As only around \SI{5}{\percent} of all recorded trips and activities are manually adjusted \cite{reporttwo}, unknown locations make up a large number of tracked trips. When looking at subsequent statistics on trip purposes, note that roughly half of all trips recorded were assigned the label \emph{unknown}. \mbox{Figure \ref{fig:MS_overview_dist}} shows the distributions of distances of trips with different purposes. Trips with purpose \emph{work/education} exhibit both the longest distance and the largest dispersion, with the \nth{3} quartile extending up to \SI{40}{\kilo\meter}. Trip types \emph{errand} and \emph{shopping} show similar distributions in regard to trip lengths. Both are relatively short with medians around \SI{4}{\kilo\meter} and smaller dispersion compared to the other purposes. A second cluster with similar distance characteristics consists of purposes \emph{home}, \emph{leisure}, and \emph{unknown}. Their medians are located around \SI{6}{\kilo\meter}. The \nth{3} quartile whisker reaches \SI{30}{\kilo\meter}.\\
A kernel density estimation of the start times of trips with different purposes is depicted in \mbox{Figure \ref{fig:MS_overview_start}}. Following Scott's rule \cite{Scott1992}, the kernel density estimation uses a Gaussian kernel with bandwith $1.8 \cdot n^{(-1/(6))}$ for $n$ samples. With regard to the temporal distribution of trip start times, a large peak for trip type \emph{work/education} occurs in the morning around 8 o'clock as to be expected. In contrast, \mbox{\emph{home}-trips} accumulate in the late afternoons and evenings. Trip types \emph{errand} and \emph{shopping} occur in two peaks each, in the morning and the afternoon with a slight dip in between. Trip type \emph{shopping} is shifted approximately one hour towards the evening. Trips with purpose \emph{leisure} increase continuously throughout the day and peak at 17 o'clock. Trips that were assigned to type \emph{unknown} do not show clear similarities to one of the aforementioned types and can thus not be reassigned on these grounds.

\subsection{Modal Split for Trip Purposes}
\begin{figure}[htp]

\sbox\twosubbox{%
  \resizebox{\dimexpr.99\textwidth-1em}{!}{%
    \includegraphics[height=3cm]{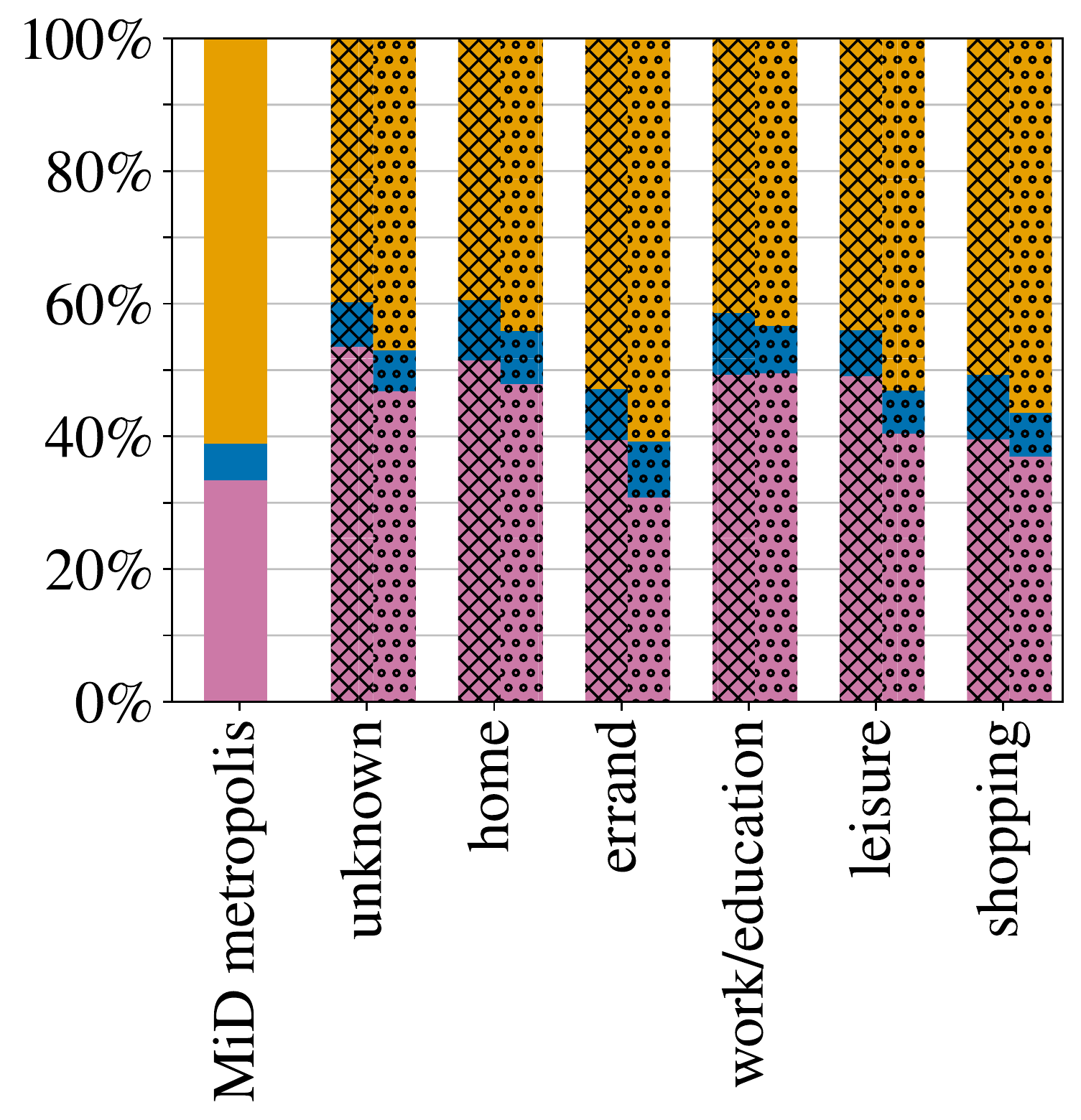}%
    \includegraphics[height=3cm]{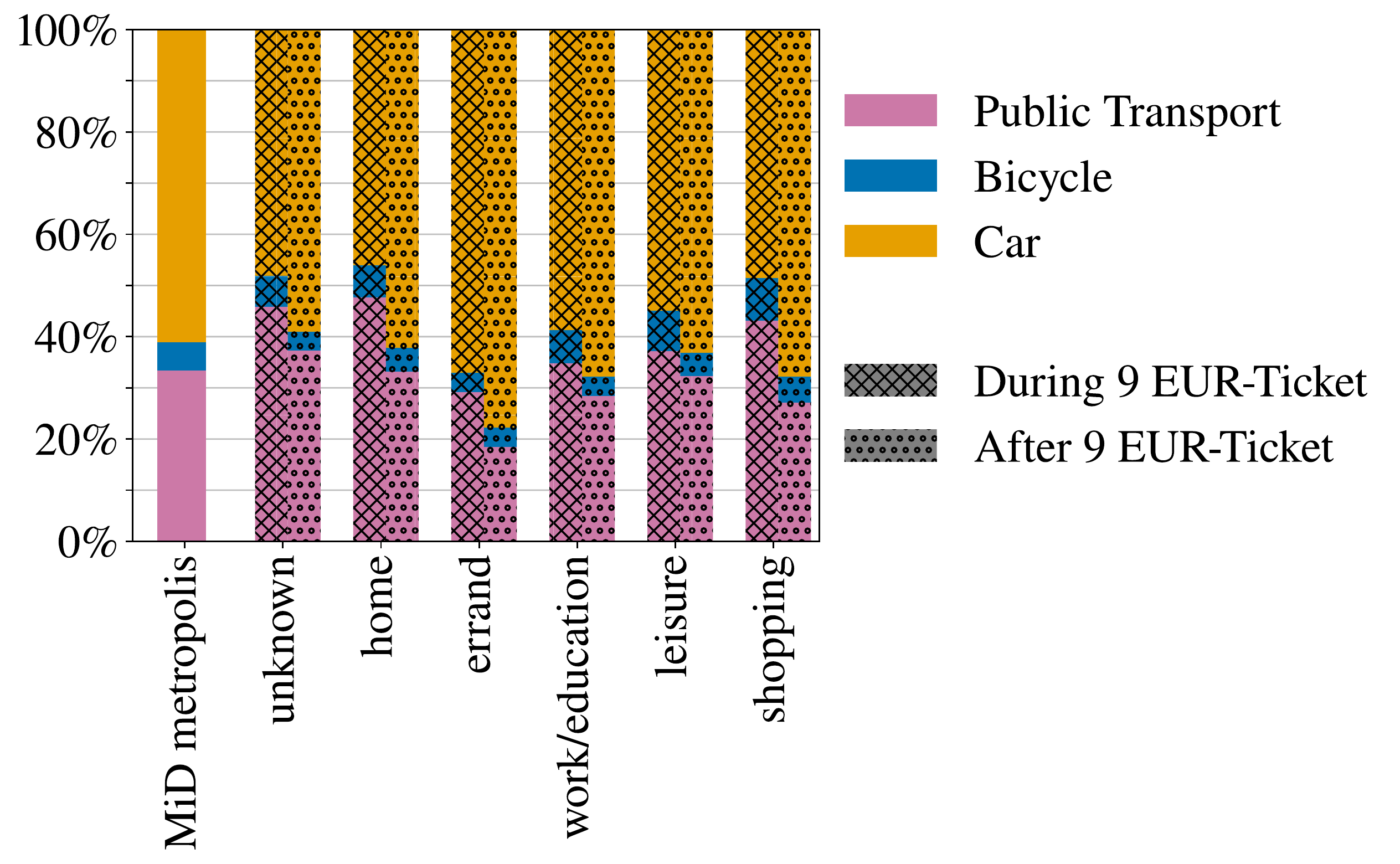}%
  }%
}
\setlength{\twosubht}{\ht\twosubbox}

\centering
\subcaptionbox{working days\label{fig:MS_purposes_wd}}{%
  \includegraphics[height=\twosubht]{_figure/purp_wd_norm_svg-tex.pdf}%
}\quad
\subcaptionbox{\mbox{non-working} days\label{fig:MS_purposes_we}}{%
  \includegraphics[height=\twosubht]{_figure/purp_we_norm_svg-tex.pdf}%
}
\caption{Modal Splits for different trip purposes, calculated using respective distances}
\end{figure}

With questioning prevailing mobility habits and routines being named one of the greatest opportunities of the \NineEuroTicket\ \footnote{\url{https://www.bundesregierung.de/breg-de/themen/deutsche-einheit/faq-9-euro-ticket-2028756}}, we examine the change in modal shares of \gls{pt}, bicycle, and car for different trip purposes in \mbox{Figures \ref{fig:MS_purposes_wd} and \ref{fig:MS_purposes_we}}. It can generally be stated that \gls{pt} usage is higher on working days than on \mbox{non-working} days with differences of \SI{5}{\percent} to \SI{10}{\percent} for respective purposes. The modal splits of our sample group are approximately similar to the one obtained from \gls{mid} on \mbox{non-working} days (with exceptions, especially \emph{errands)} but significantly higher on working days. Trips of purpose \emph{errands} yield a substantially smaller share of \gls{pt} and a larger share of car respectively, compared to the other purposes both for working and \mbox{non-working} days. For all but one purpose, the share of \gls{pt} significantly increased during the \NineEuroTicket\ period. Purpose \emph{work/education} during the week is the only group where no clear increase in \gls{pt} is evident. The decreases in \gls{pt} ridership range from single digits to \SI{10}{\percent}. The share of cycling is mostly constant, meaning most trips not conducted with \gls{pt} after the \NineEuroTicket\ period were replaced by car. The largest drop in \gls{pt} ridership can be observed for trip purpose \emph{leisure} on working days. On \mbox{non-working} days, type \emph{shopping}'s \gls{pt} ridership dropped the most.

\subsection{Modal split by income group}
\begin{figure}[htp]

\sbox\twosubbox{%
  \resizebox{\dimexpr.99\textwidth-1em}{!}{%
    \includegraphics[height=3cm]{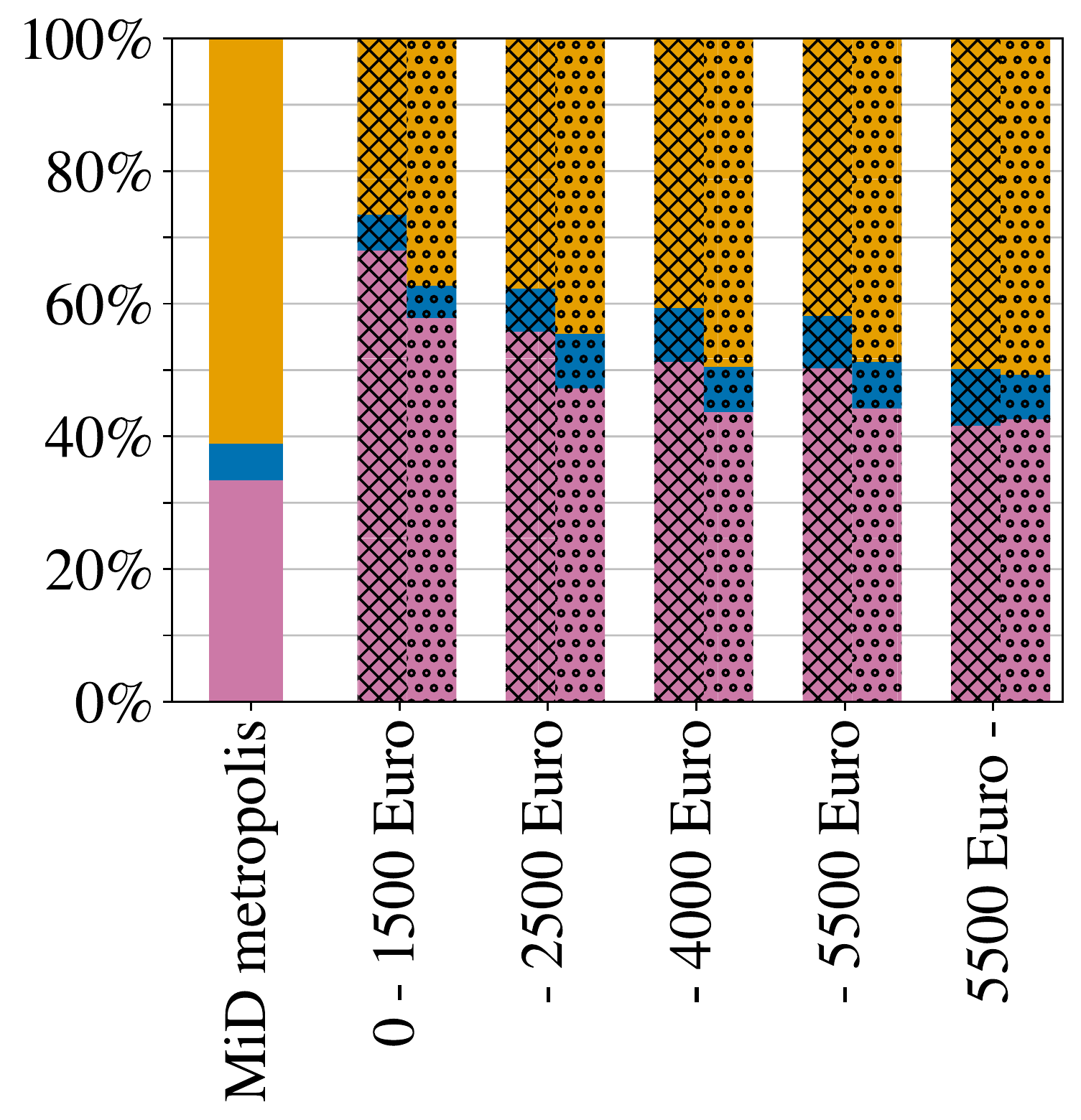}%
    \includegraphics[height=3cm]{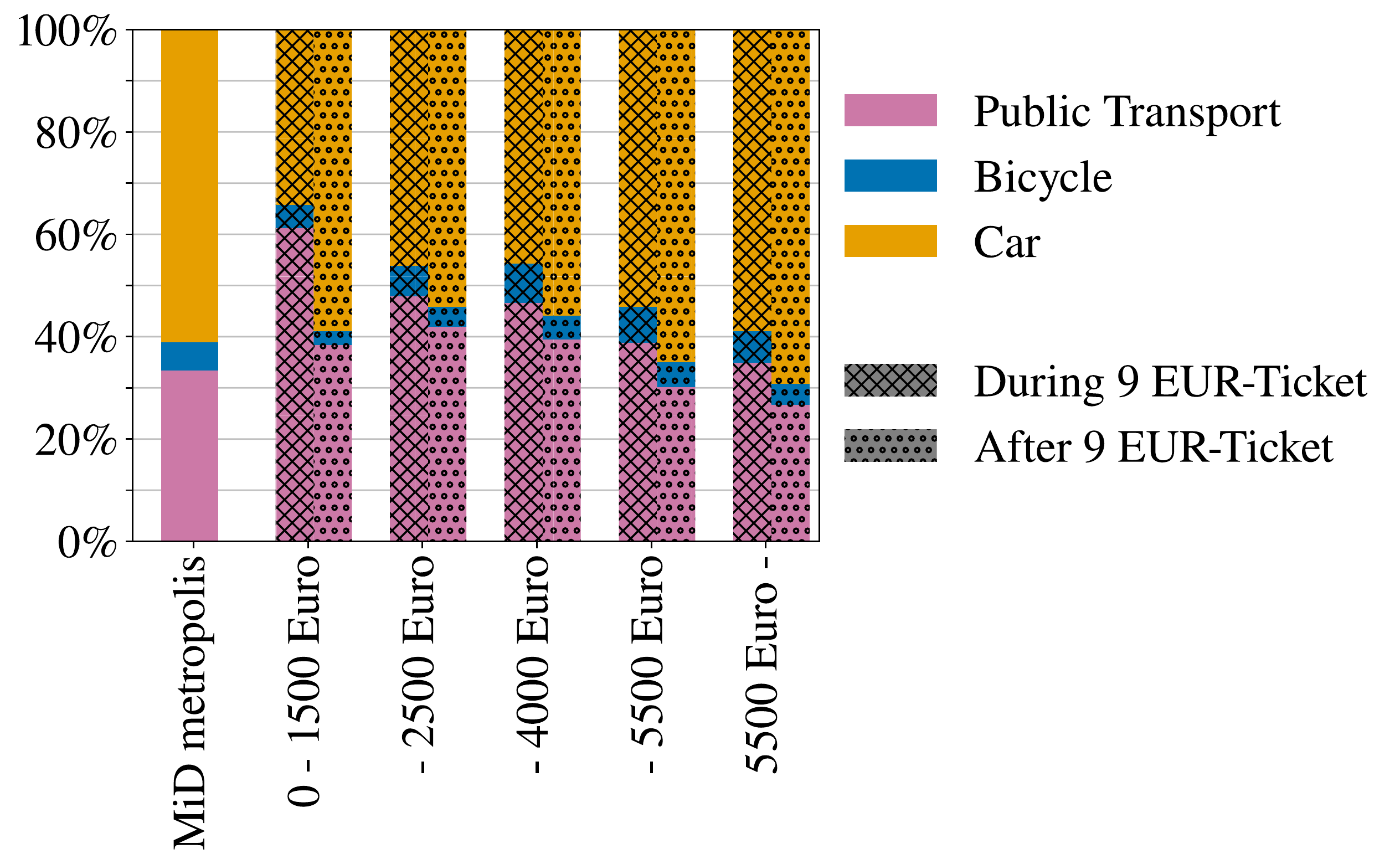}%
  }%
}
\setlength{\twosubht}{\ht\twosubbox}

\centering
\subcaptionbox{working days\label{fig:MS_income_groups_wd}}{%
  \includegraphics[height=\twosubht]{_figure/inc_wd_norm_svg-tex.pdf}%
}\quad
\subcaptionbox{\mbox{non-working} days\label{fig:MS_income_groups_we}}{%
  \includegraphics[height=\twosubht]{_figure/inc_we_norm_svg-tex.pdf}%
}
\caption{Modal Splits for different income groups, calculated using respective distances}
\end{figure}

The second major goal of the \NineEuroTicket\ was to relieve the burden on households whose cost of living had risen sharply as a result of current price developments, which especially impacts households with lower incomes. We therefore investigate the impact of the ticket on different income groups in \mbox{Figures \ref{fig:MS_income_groups_wd} and \ref{fig:MS_income_groups_we}}.\\
For all income groups, \gls{pt} usage is higher during the week than on weekends and generally exceed the one from \gls{mid}. The results of our sample group illustrate how \gls{pt} ridership generally decreases inversely proportional to income. Furthermore, the desired effect of the \NineEuroTicket\ (increased \gls{pt} ridership) on working days likewise relates inversely proportional to income. On \mbox{non-working} days, the picture is different. The largest drop in \gls{pt} usage of approximately 20 \% is to be observed for the group with the lowest income while the other groups show similar absolute decreases, although from different starting points. The trips were entirely replaced by motorized individual transport. The share of cycling is lowest for income group \mbox{\emph{0 - 1500€}}, especially on working days. With one exception, \gls{pt} ridership increased on all occasions during the period of the \NineEuroTicket. This exception is posed by the income group \mbox{\emph{more than 5,500 €}}, where the level stayed constant. While the share of cycling did not change significantly between phases, income group \mbox{\emph{1,500€ - 2,500€}} increased their share of cycling on working days after the end of the \NineEuroTicket\ and replaced a significant amount of trips conducted wit \gls{pt} before.

\subsection{Modal Split of Age Groups}

\begin{figure}[htp]
\sbox\twosubbox{%
  \resizebox{\dimexpr.99\textwidth-1em}{!}{%
    \includegraphics[height=3cm]{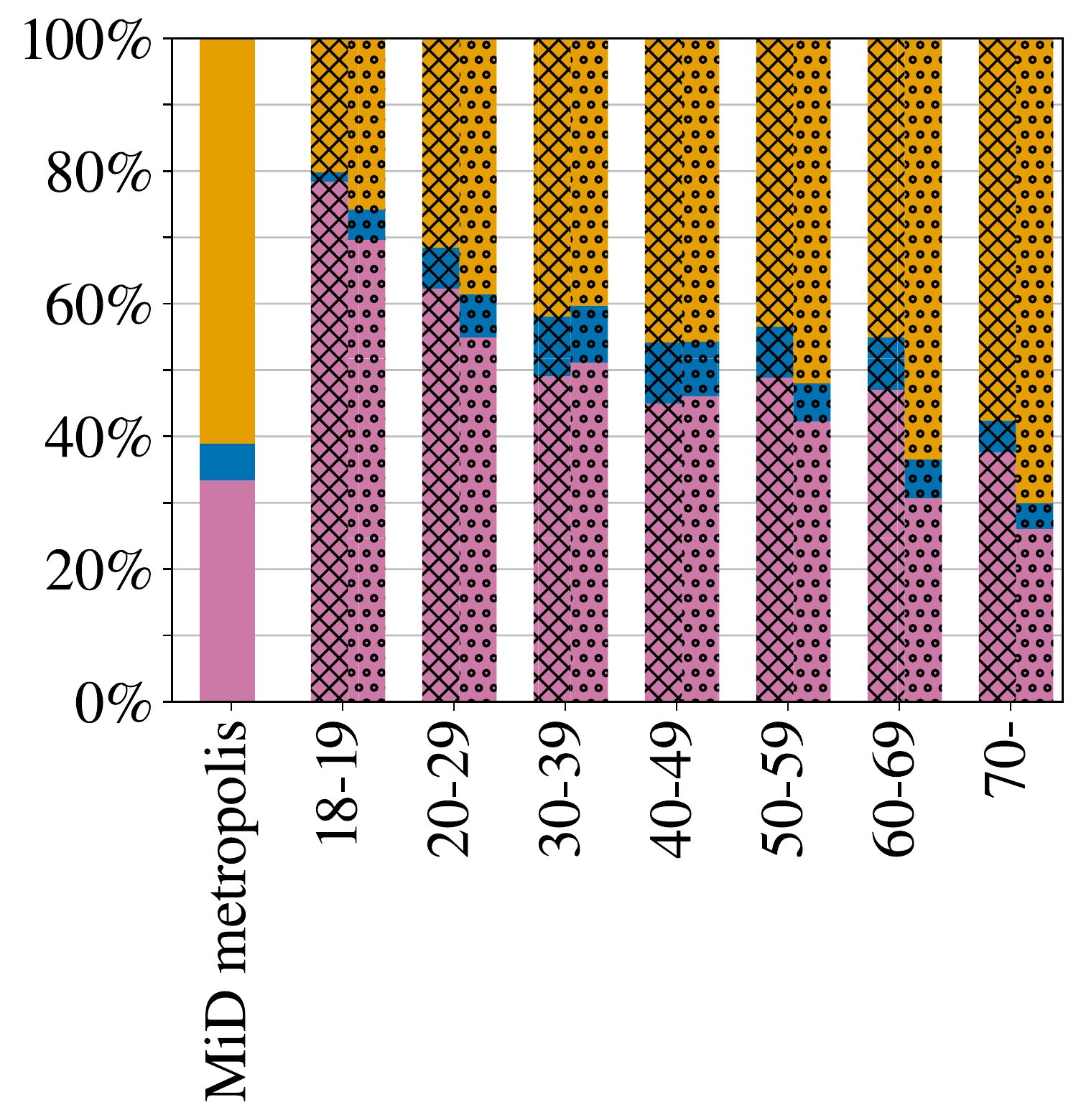}%
    \includegraphics[height=3cm]{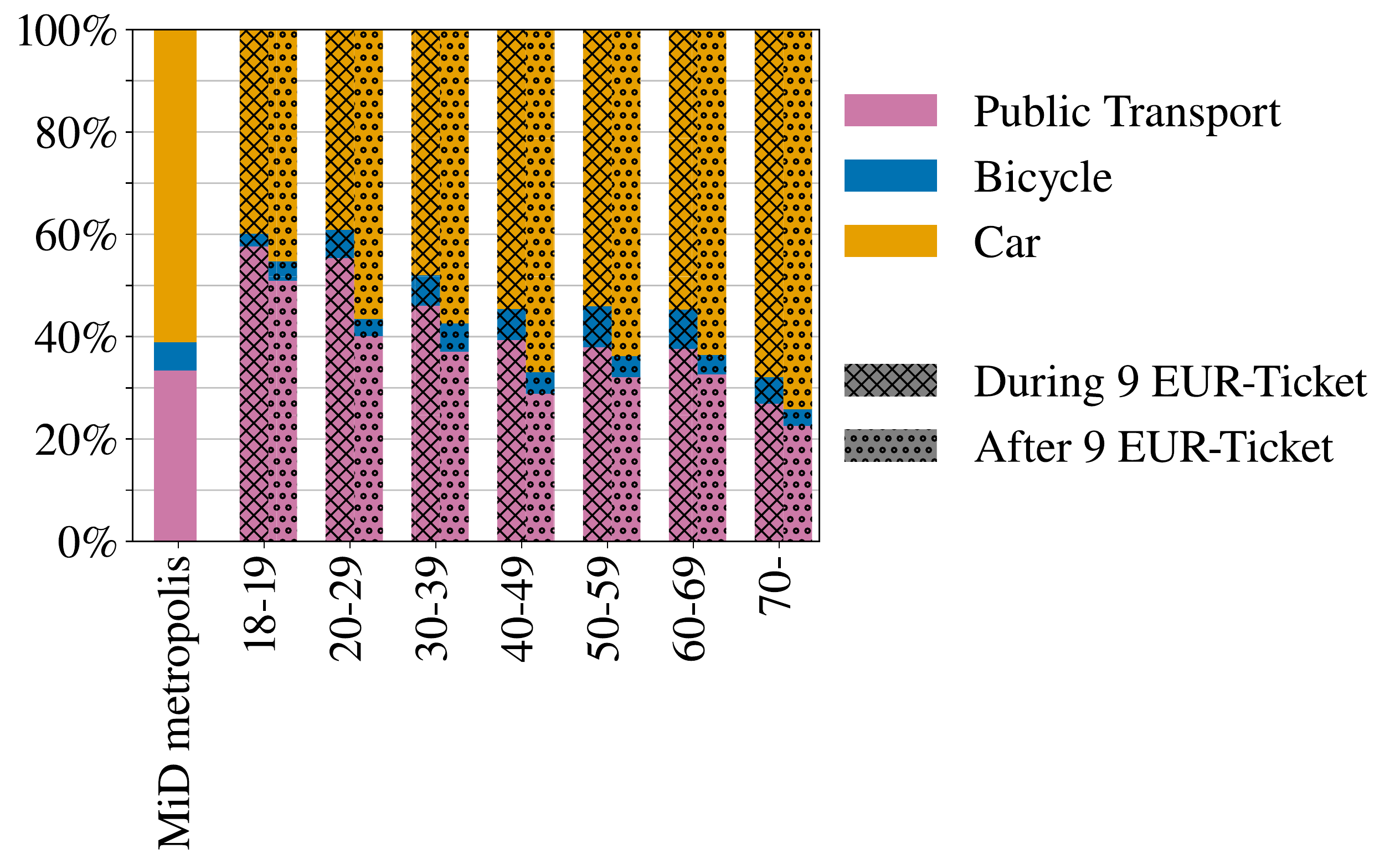}%
  }%
}
\setlength{\twosubht}{\ht\twosubbox}

\centering
\subcaptionbox{working days\label{fig:MS_age_groups_wd}}{%
  \includegraphics[height=\twosubht]{_figure/age_wd_norm_svg-tex.pdf}%
}\quad
\subcaptionbox{\mbox{non-working} days\label{fig:MS_age_groups_we}}{%
  \includegraphics[height=\twosubht]{_figure/age_we_norm_svg-tex.pdf}%
}
\caption{Modal Splits for different age groups, calculated using respective distances}
\end{figure}

Concluding the analysis of the \NineEuroTicket's impact on demographic groups, we explore the effects on the modal splits of different age groups in \mbox{Figures \ref{fig:MS_age_groups_wd} and \ref{fig:MS_age_groups_we}}. The data of our sample group reveals a general trend of decreasing/increasing \gls{pt}/car ridership, respectively, proportionally to age. This trend is stronger on working days, while the modal split of age groups between 40 and 70 years on \mbox{non-working} days is approximately similar. Another finding valid for all age groups is a higher \gls{pt} usage on working days than on \mbox{non-working} days. Compared to \gls{mid}, the share of \gls{pt} is higher except age groups above 40 years on \mbox{non-working} days. Distances traveled by bicycle are comparable, although the youngest age group distinctively exhibits the smallest relative share of active mobility on both kinds of days.\\
On working days, the age groups \emph{30-39} and \emph{40-49} both yield higher shares of \gls{pt} usage after the \NineEuroTicket\, although the absolute increase is small. The opposite is true for all other age groups on working days. On \mbox{non-working}days, the only group showing a higher \gls{pt} usage after the period of the \NineEuroTicket\ is the one below 20 years of age. The largest drop of \gls{pt} usage is to be found for the 20 to 29 year old participants on \mbox{non-working} days. Further comparably large decreases are evident for the oldest two age groups above 60 years, on working as well as \mbox{non-working} days, although a large share of participants is likely to be retired.

%% file: _sections/5_discussion.tex
\newpage
\section{Discussion and Conclusion}
\label{sec:discussion}

This report presents first analyses of the data collected within the first four months of the tracking experiment of the ''Mobilität.Leben'' study. As a baseline for the understanding and interpretation of all subsequent analyses, it assesses age, gender, income, and dwellings of the tracking participants. Subsequently, general transport statistics, as well as disaggregated statistics on activities and the usage of different means of transport are presented with a special focus on modal splits. This section discusses our findings and the limitations of both the study and the employed methodology.\\
A first limitation of the study is constituted by the size of our sample group as well as the available \mbox{time-frame} for the assessment of \mbox{pre-\NineEuroTicket} behaviors. While participation rates have shown to be constantly high throughout the reported \mbox{time-frame}, only one week of data before the introduction of the \NineEuroTicket\ exists. In addition, due to the successive \mbox{start-up} phase, far fewer participants submitted data during this first week than in later phases of the experiment. Seasonal effects may also be present during all phases of the experiment since all data was recorded in less than a year's time. This lack of data, especially in the first phase of our experiment, motivated the decision to leave out any data recorded before the start of the \NineEuroTicket\ for more disaggregated analyses. In consequence for purpose- or \mbox{subgroup-specific} statistics, there is no direct reference from before the \NineEuroTicket\ from within the study group. When analyzing the impact of the \NineEuroTicket\ by \mbox{e.g.} a comparison of modal splits \textit{during} and \textit{after} its introduction, potential backlashes from \mbox{re-increased} pricing or other compensatory mechanisms may also influence mode choice without any indication in the presented data. Although the \gls{mid} survey can serve as an alternative reference for \mbox{pre-\NineEuroTicket} behaviors to some extent, large differences between \gls{mid} and our surveys can be observed especially for modal splits. These differences appear to be in the same order of magnitude as the observed influences of the \NineEuroTicket\ indicating that no clear baseline of mobility behavior prior to the \NineEuroTicket\ can be established. Considering the differences in survey methodology of \gls{mid} and ''Mobilität.Leben'' supports this argument. \\
Apart from the particularities concerning the first phase of the experiment, it should be noted that all the results presented in this report are initially only representative of the group of participants observed. This group is \textendash\ as shown in \mbox{Section \ref{sec:tracking_participants}} \textendash\ relatively wealthy, a little younger than the German population, and includes more male than female participants. Whether and to which extent observations regarding this study group can be extrapolated onto larger populations in and outside of Germany remains unclear.\\ 
Another point worthy of discussion is the way trip purposes and means of transport are labeled, which are \textendash\ for the best part \textendash\ not surveyed directly, but inferred from raw GPS trajectories supplied by the participants. This is done based on the algorithms employed by the MOTIONTAG smartphone app. While tracking participants have the option to correct any labeling manually and thereby directly reveal activity types and mode choice, only around \SI{5}{\percent} of all recorded trips and activities are manually adjusted \cite{reporttwo}. Approximately half of all recorded activities retain an \textit{unknown} label.\\  
Despite all limitations and reservations due to sample size, quality, and sample group composition, a clear positive effect of the \NineEuroTicket\ on the aggregated modal share on \gls{pt} usage is revealed by our experiment for both working days and \mbox{non-working} days. \\
The disaggregated modal splits shown in \mbox{Section \ref{sec:data_analysis}} have revealed significant influences of trip purpose, income, and age on the modal split and \gls{pt} utilization respectively. Most notably, work and education trips during workdays were hardly affected by the \NineEuroTicket. In contrast, errands, leisure, and shopping activities were preceded by \gls{pt} trips more often during the \NineEuroTicket\ program than afterwards. Independent of the experiment phase, increasing household incomes correlate with decreasing \gls{pt} usage and increased car usage in our sample. Additionally, the availability of a reduced ticket price for \gls{pt} appears to affect \mbox{low-income} households significantly stronger than their \mbox{high-income} counterparts especially on \mbox{non-working} days. Similarly, age presents to have an adverse effect on \gls{pt} usage in our sample. Nevertheless, the relative reduction of \gls{pt} utilization after the end of the \NineEuroTicket\ is especially large for participants under 30 and over 60 years of age.